\documentclass[pra,showpacs,showkeys]{revtex4}% Physical Review AI

\usepackage{graphicx}% Include figure files
\usepackage{dcolumn}% Align table columns on decimal point
\usepackage{bm}% bold math
\begin{document}

\title{Nonlinear mechanisms to Rogue events in the process of  interaction between  optical filaments}

\author{L. M. Kovachev$^{1}$ and D. A. Georgieva $^{2}$}
\affiliation{ $^{1}$Institute of Electronics, Bulgarian Academy of Sciences, \\ 72 Tzarigradcko shossee, 1784 Sofia, Bulgaria;\\
 $^{2}$Faculty of Applied Mathematics and Computer Science,
Technical University of Sofia, 8 Kliment Ohridski Blvd., 1000 Sofia,
Bulgaria}

\date{\today}

\begin{abstract}
We investigate two types of nonlinear interaction between collinear
femtosecond laser pulses with power slightly above the critical for
self-focusing $P_{cr}$.  In the first  case we study energy exchange
between filaments. The model describes this  process  through
degenerate four-photon parametric mixing (FPPM) scheme and requests
initial phase difference between the waves. When there are no
initial phase difference between the pulses, the FPPM process does
not work. In this case it is obtained the  second type of
interaction as merging between two, three or four filaments in a
single filament with higher power.  It is found that in the second
case the interflow between the  filaments  has potential  of
interaction due to   cross-phase modulation (CPM).
\end{abstract}


\maketitle

\section{Introduction}

When two, three, four or higher number of filaments propagate in
parallel and close trajectories, interflow and observation of
extreme events are  reported  \cite{TZ, KOZ,CHEK, SKUP, WAH}.  These
mergers, with appearing of a strong filament, were called Rogue
events during the filamentation process. The coupling between two
filaments was experimentally observed for first time in \cite{TZ}.
The conditions for optimal coupling were obtained later from
Kosareva et al. \cite{KOZ} and  the optimum  is  for filaments with
power near to critical for self-focusing and small diameter of the
spot $r_0<3mm$.  As it was reported recently in \cite{GAETA},
depending on the relative phase and the incidence angle, the
filament can experience fusion, repulsion, energy redistribution and
spirial motion. The  experiments in xenon gas were performed with
several gigawatts input peak power \cite{SKUP} to obtain parallel
filament strings with numbers $N <12$. The laser pulse breaks up
from spatially homogeneous beam profile into several highly
localized filament strings each with pulse power slightly above the
critical for self-focusing $P_{cr}$ \cite{BERGE, BOYD}. This is the
reason to look for nonlinear optical mechanisms leading to exchange
of energy or mergers during the process of multifilament
propagation. The three dimensional localization appears similar to
the soliton interaction  in one-dimensional system as optical
fibers, and based on clamping effects due to CPM \cite{MEN1, MEN2,
KOV1, KOV2} and FPPM \cite{KOV2}.  The nonlinear interaction process
in fibers strongly depends on the initial phase difference between
the pulses.

In this paper we propose a nonlinear vector model,  where in details
is investigated the role of CPM and degenerate FPPM processes in
respect to the relative movements of laser filaments. We investigate
numerically the interaction between optical pulses in the cases
when: 1) the initial phase difference between pulses  is not equal
to zero $\Delta\varphi\neq 0$ and 2) the initial optical pulses
admit equal phases $\Delta\varphi=0$. Thus, by properly selected
initial conditions, we take into account  the FPPM process as
addition to the CPM influence. The proposed nonlinear vector model
is investigated numerically on the base of the split-step Fourier
method. We introduce by the moment formalism nonlinear acceleration
and potentials between the weight centrums of the pulses.

\section{Theory: Nonlinear Polarization and basic system of equations}

As it was pointed in \cite{KOL1, KOL2, KOVBOOK}, the filamentation
process can be described more correctly by using the generalized
nonlinear operator

\begin{eqnarray}
\label{NLTH} \vec{P}^{nl} = n_2 \left( \vec{E} \cdot \vec{E}
\right)\vec{E},
\end{eqnarray}
which includes additional processes associated with third harmonic
generation. We substitute into the nonlinear operator (\ref{NLTH})
two-component electrical vector  $\vec{E} = (E_x, E_y, 0)$ at one
carrying frequency $\omega_0$

\begin{eqnarray}
\label{ELF} \vec{E} =\frac{ \left( A_x \exp\left[i(\omega_0 t-k_0z)\right] + c.c.
\right)}{2}\vec{x} + \frac{ \left( A_y \exp\left[i(\omega_0 t-k_0z)\right] + c.c.
\right)}{2}\vec{y},
\end{eqnarray}
where $A_x=A_x(x, y, z, t), A_y = A_y(x, y, z, t)$ are the amplitude
functions and  $k_0$ is the carrying wave number of the laser
source.

The nonlinear polarization (\ref{NLTH}) generates the following components

\begin{eqnarray}
\label{POLTH} \vec{P}^{nl}_x = \tilde{n}_2 \left[ \frac{1}{3} \left(
A_x^2 + A_y^2\right) A_x \exp\left[2i(\omega_0 t-k_0z)\right] +
\left( |A_x|^2 + \frac{2}{3}|A_y|^2\right) A_x + \frac{1}{3} A_x^*
A_y^2
\right]\exp\left[i(\omega_0 t-k_0z)\right] + c.c. \nonumber\\
\\
\vec{P}^{nl}_y =  \tilde{n}_2 \left[ \frac{1}{3} \left( A_x^2 +
A_y^2\right) A_y \exp\left[2i(\omega_0 t-k_0z)\right] + \left(
|A_y|^2 + \frac{2}{3}|A_x|^2\right) A_y + \frac{1}{3} A_y^* A_x^2
\right]\exp\left[i(\omega_0 t-k_0z)\right] + c.c., \nonumber
\end{eqnarray}
where $\tilde{n}_2=\frac{3}{8}n_2$. The operator $ n_2 \left(
\vec{E} \cdot \vec{E} \right)\vec{E}$ generalizes the case of Marker
and Terhune's operator, and includes to the self-action terms, CPM
terms, FPPM terms and also additional terms associated with
Third-Harmonic Generation  (THG).

The initial laser pulses $\left( t_0\geq 50 fs \right) $ possess a
relatively narrow-band spectrum $\left(\Delta k_z \ll k_0 \right)$
($\Delta k_z$ is the spectral pulse width). During the filamentation
process the initial self-focusing broadens significantly the pulse
spectrum. The broad-band spectrum $\left(\Delta k_z \sim k_0
\right)$ is one of the basic characteristics of the stable filament.
The dynamics of broad-band pulses can be presented properly within
different non-paraxial models  such as UPPE  \cite{KOL1, KOL2} or
non-paraxial envelope equations \cite{KOVBOOK}. Another standard
restriction in the filamentation theory is the use of one-component
scalar approximation of the electrical field $\vec{E}$. This
approximation though,  is in contradiction with recent experimental
results, where rotation of the polarization vector is observed
\cite{KOSA,ZIG}. For this reason, in the present paper we use the
non-paraxial vector model up to second order of dispersion, in which
the nonlinear effects are described by the nonlinear polarization
components (\ref{POLTH}). The system of non-paraxial equations of
the amplitude functions $A_x, A_y$ of the two-component electrical
field (\ref{ELF}) has the form

\begin{eqnarray}
\label{SYSGAL} -2i\frac{k_0}{v_{gr}} \frac{\partial A_x}{\partial
t}= \Delta_{\bot} A_x  - \frac{\beta + 1}{v_{gr}} \left(
\frac{\partial^2 A_x}{\partial t^2} - 2v_{gr} \frac{\partial^2
A_x}{\partial t
\partial z}\right) - \beta \frac{\partial^2 A_x}{\partial z^2}
\nonumber\\
+ k^2_0 \tilde{n}_2 \left[\frac{1}{3} \left(A^2_x + A^2_y
\right)A_x\exp\left(2ik_0\left(z-\left(v_{ph}-v_{gr}\right)t\right)\right)
+ \left( |A_x|^2 + \frac{2}{3}|A_y|^2\right) A_x + \frac{1}{3}
A^*_xA^2_y \right] \nonumber\\
\\
-2i\frac{k_0}{v_{gr}} \frac{\partial A_y}{\partial t}= \Delta_{\bot}
A_y - \frac{\beta + 1}{v_{gr}} \left( \frac{\partial^2 A_y}{\partial
t^2} - 2v_{gr} \frac{\partial^2 A_y}{\partial t
\partial z}\right) - \beta \frac{\partial^2 A_y}{\partial z^2}
\nonumber\\
+ k^2_0 \tilde{n}_2 \left[\frac{1}{3} \left(A^2_x + A^2_y
\right)A_y\exp\left(2ik_0\left(z-\left(v_{ph}-v_{gr}\right)t\right)\right)
+ \left( |A_y|^2 + \frac{2}{3}|A_x|^2\right) A_y+ \frac{1}{3}
A^*_yA^2_x \right], \nonumber
\end{eqnarray}
where  $v_{gr}$ and $v_{ph}$ are the
group and phase velocities correspondingly, $\beta = k_0v_{gr}^2k''$
and $k''$ is the group velocity dispersion.

\begin{figure}[b]
\begin{center}
\begin{tabular}{c}
\includegraphics[width=130mm,height=70mm]{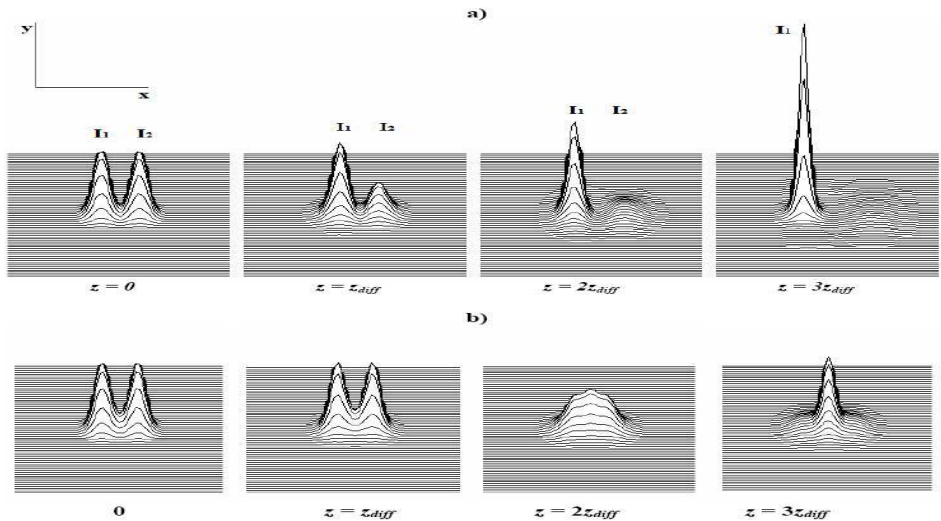}
\end{tabular}
\end{center}
\caption{ (1a) Energy exchange between two collinear filaments
$\vec{A}_1$ and $\vec{A}_2$ with power slightly above the critical
$P_{cr}$ ($\gamma=1.5$). The pulses are separated initially at
distance $2a=3.4$ and the evolution is governed by the system of
equations (\ref{SYSGAL1}). In the initial conditions (\ref{INCOND1})
the phase difference $\varphi=\pi/4$ correspond to maximal energy
exchange. Due to degenerated FPPM process one of the filaments is
amplified while the other filament enters in linear mode and
vanishes.  (1b) Fusion between the same pulses when they are with
equal initial phases, i. e. $\Delta\varphi=0$. The similar picture
is seen when the FPPM terms are excluded from the equations
(\ref{SYSGAL1}) and only interaction due to CPM is investigated.
With $z_{diff}$ is denoted the diffraction length
$z_{diff}=k_0r_0^2$.}
\end{figure}

This model describes the ionization-free filamentation regime, where
the pulse intensities are close to the critical one for
self-focusing. The first nonlinear term in (\ref{SYSGAL})
corresponds to  coherent GHz generation \cite{KOVBOOK}. The system
(\ref{SYSGAL}) is written in Galilean frame $\left( z'=z-vt; t'=t
\right)$.  In all  coordinate systems - laboratory, moving in time,
and Galilean, the group velocity adds an additional phase
(carrier-envelope phase) in the third harmonic terms and transforms
them to GHz ones. This can be seen directly for the system
(\ref{SYSGAL}) written in Galilean frame, which determines the
choice of coordinates. The last nonlinear term in (\ref{SYSGAL})
describes degenerate four-photon parametric mixing. To satisfy the
Manley-Rowe relations of the truncated equations with a generalized
nonlinear polarization of the type $\vec{P}^{nl} = n_2 \left(
\vec{E} \cdot \vec{E} \right)\vec{E}$,  some restrictions on the
components of the electrical field are imposed. The condition is
simple - the possible initial components $A_x$ and $A_y$ should be
complex-conjugated fields. The conservation laws give us additional
information on the behavior of the vector amplitude function: \emph{
only components of the vector amplitude field $\vec{A}=(A_x,A_y,0)$,
which present rotation of the vector $\vec{A}$ in the plane $(x,y)$,
satisfy the MR conditions}. That is why in our numerical
experiments, as well as in our analytical investigations, we will
use complex-conjugated components only.

The system of equations (\ref{SYSGAL}) written in dimensionless form
becomes
\begin{eqnarray}
\label{SYSGAL1} -2i\alpha\delta^2 \frac{\partial A_x}{\partial t}=
\Delta_{\bot} A_x  - \delta^2\left(\beta + 1\right) \left(
\frac{\partial^2 A_x}{\partial t^2} - \frac{\partial^2 A_x}{\partial
t \partial z}\right) - \delta^2\beta \frac{\partial^2 A_x}{\partial
z^2}
\nonumber\\
+ \gamma \left[\frac{1}{3} \left(A^2_x + A^2_y
\right)A_x\exp\left(2i\alpha\left( z-\Delta\tilde{
v}_{nl}t\right)\right) + \left( |A_x|^2 + \frac{2}{3}|A_y|^2\right)
A_x + \frac{1}{3}
A^*_xA^2_y \right] \nonumber\\
\\
-2i\alpha\delta^2 \frac{\partial A_y}{\partial t}= \Delta_{\bot} A_y
- \delta^2\left(\beta + 1\right) \left( \frac{\partial^2
A_y}{\partial t^2} - \frac{\partial^2 A_y}{\partial t
\partial z}\right) - \delta^2\beta \frac{\partial^2 A_y}{\partial z^2}
\nonumber\\
+ \gamma \left[\frac{1}{3} \left(A^2_x + A^2_y
\right)A_y\exp\left(2i\alpha\left( z-\Delta\tilde{
v}_{nl}t\right)\right) + \left( |A_y|^2 + \frac{2}{3}|A_x|^2\right)
A_y + \frac{1}{3} A^*_yA^2_x \right], \nonumber
\end{eqnarray}
where $x = x/r_0$, $y = y/r_0$, $z = z/r_0$ are the dimensionless
coordinates, $r_0$ is the pulse waist, $z_0 = v_{gr}t_0$ is the
spatial pulse length, $\alpha=k_0z_0$, $\delta = r_0/z_0$, $\gamma =
k^2_0r^2_0\tilde{n}_2|A_0|^2/2$ is the nonlinear coefficient and
$\Delta\tilde{v}_{nl} = \left(v_{ph}-v_{gr}\right)/v_{gr}$ is the
normalized group-phase velocity difference.

\section{Numerical simulations}

In the experiments on multi-filamentation two basic trends are
observed. The first one is that  the number of filaments is reduced
significantly as a function of the distance \cite{KILO}. The second
trend is observed recently in \cite{TZ, KOZ,CHEK, SKUP, WAH} as
mergers between two, tree or four filaments in one  (Rogue) wave. We
think that the both processes are connected and they are results of
different types of nonlinear interaction. Therefore  we investigated
in details interaction of two filaments. By control of the initial
phase difference between the pulses it is possible  to include or
exclude the process of FPPM in the nonlinear interaction. When the
initial phase difference between the pulses is not equal to zero,
the process of  FPPM starts to work and an intensive exchange of
energy is observed \cite{DAN}. When the initial phase difference of
the pulses is equal to zero the process of  FPPM practically does
not work and the nonlinear interaction is due the CPM process.

The numerical results are presented for initial conditions:  $120
fs$ Gaussian bullets with waist and spatial length $r_0 = z_0 = 72
\mu m $ and power slightly above $P_{cr}$. In this case $\alpha =
90\pi$, $\delta = 1$,  $\Delta\tilde{v}_{nl}=0.00023$ and $\gamma
\in 1.5-3$. The phase difference between the $A_x$ and $A_y$
components is initially $\pi/2$ in order to satisfy the conservation
laws. We investigate two collinear laser pulses as two vector fields
$\vec{A}_1$ and $\vec{A}_2$ at small distance $a$  between them.
Each of the pulses admits $\vec{x}$ and $\vec{y}$ components:
$\vec{A}_j =A_{j,x} \vec{x} + A_{j,y} \vec{y}, j = 1,2$. The initial
conditions for numerical solution of the system of equations
(\ref{SYSGAL1}) have the form

\begin{eqnarray}
\label{INCOND1}  A_x = A_{1,x} + A_{2,x} = \frac
{A^0_1}{\sqrt{2}}\exp\left(-\frac{(x+a)^2 + y^2 +
z^2}{2}\right) \nonumber \\
+ \frac{A^0_2}{\sqrt{2}}\exp\left(-\frac{(x-a)^2 + y^2
+z^2}{2}\right)
\exp\left(i\Delta\varphi\right) \nonumber\\
\\
A_y = A_{1,y} + A_{2,y} =  \Biggr \{ \frac
{A^0_1}{\sqrt{2}}\exp\left(-\frac{(x+a)^2 + y^2 +
z^2}{2}\right) \nonumber \\
+\frac{A^0_2}{\sqrt{2}}\exp\left(-\frac{(x-a)^2 + y^2
+z^2}{2}\right)
\exp\left(i\Delta\varphi\right) \Biggr \}
\exp\left(i\frac{\pi}{2}\right), \nonumber
\end{eqnarray}
where $A_x$ and $A_y$ are composed of the $x$- and $y$-components of
the two optical pulses propagating along different parallel
trajectories. The phase difference between the $A_x$ and $A_y$
components is initially $\pi/2$ in order to satisfy the conservation
laws, while the phase difference between the pulses is denoted by
$\Delta\varphi$. By varying the phase difference $\Delta\varphi$ we
include (and exclude, when  $\Delta\varphi=0$) the FPPM process. The
interaction of optical pulses $\vec{A}_1$ and $\vec{A}_2$ for
$\gamma = 1.5$, $\Delta v = 1.5$, $2a = 3.4$ and
$\Delta\varphi=\pi/4$ is shown on Fig. 1a. It is observed that  the
amplified pulse self-focuses and gets enough power to continue its
propagation, while the other pulse gives out energy, enters into
linear mode and vanishes. In this way the number of filaments can be
reduced by non-linear parametric processes in $\chi^{(3)}$ media. In
the following numerical experiment (Fig. 1b) we exclude the FPPM
process by using initial phase difference $\Delta\varphi=0$. To
verify this result  we also exclude the parametric step from the the
system of equations (\ref{SYSGAL1}) and increase the intensity by
factor $1/4$ to keep on  the critical power.  The pulses start to
attract each other without energy exchange and as result a merging
and  self-focusing  are observed. In the both numerical experiments
(with $\Delta\varphi=0$ or when the parametric step is excluded from
the program) the results are similar - there is no energy exchange
and the fusing between the filaments is clearly seen. Similar
potential type of interaction by CPM was reported in optical fibers
\cite{KOV1, KOV2}. In the following section of this paper we
calculate the nonlinear acceleration and potential between the
weight centers of optical pulses. In the general case of few optical
filaments usually there are random phase differences between the
waves. On Fig. 2  the interaction between three pulses  is
presented. Two of them  are with equal initial phases, while the
third one admits phase difference $\Delta\varphi=\pi/4$  in respect
to others. Similar dependance on the initial phase difference is
observed: fusion between the pulses with equal phases, while the
third one exchanges energy by FPPM process.

\begin{figure}
\begin{center}
\begin{tabular}{c}
\includegraphics[width=140mm,height=55mm]{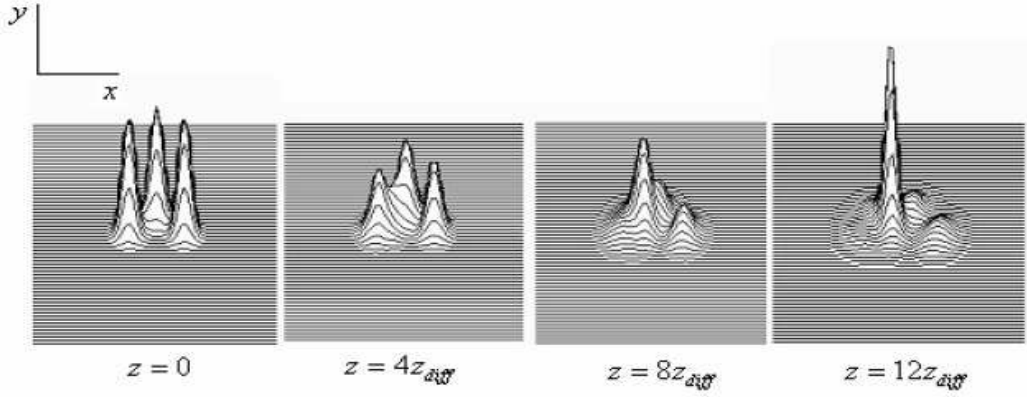}
\end{tabular}
\end{center}
\caption{Interaction between three collinear filaments $\vec{A}_1$,
$\vec{A}_2$ and $\vec{A}_3$ governed by the system of equations
(\ref{SYSGAL1}). Two of the pulses  are with equal initial phases,
while the third admits phase difference $\Delta\varphi=\pi/4$  in
respect to others. Fusing between the pulses with equal phases is
seen, while the third one exchanges  energy by FPPM process.}
\end{figure}

\section{Moment formalism and potentials}

To obtain analytical expressions of the influence of CPM on the
relative moving of optical pulses we exclude   the FPPM process  and
GHz generation from the system of equations (\ref{SYSGAL1}). The
basic system in this case is transformed to $(3+1)D$  of Manakov
type

\begin{eqnarray}
\label{SYS1} -2ik_0\left[\frac{1}{v_{gr}} \frac{\partial
A_x}{\partial t}+\frac{\partial A_x}{\partial z}\right]= \Delta
A_x - \frac{1+k_0 v_{gr}^2 k^{''}}{v_{gr}^2}  \frac{\partial^2
A_x}{\partial t^2} + k^2_0 \tilde{n}_2 \left( |A_x|^2 +
\frac{2}{3}|A_y|^2\right) A_x \nonumber\\
\\
-2ik_0\left[\frac{1}{v_{gr}} \frac{\partial A_y}{\partial
t}+\frac{\partial A_y}{\partial z}\right]= \Delta A_y -
\frac{1+k_0 v_{gr}^2 k^{''}}{v_{gr}^2} \frac{\partial^2
A_y}{\partial t^2} + k^2_0 \tilde{n}_2 \left( |A_y|^2 +
\frac{2}{3}|A_x|^2\right) A_y \nonumber.
\end{eqnarray}
The scalar case of nonlinear interaction is investigated in
\cite{KOV2}. In this paper we will preform similar analysis applied
to collinear laser pulses presented as vector fields $\vec{A}_1$ and
$\vec{A}_2$. We decompose as in the previous section the vectors
$\vec{A}_1$ and $\vec{A}_2$ in $(x, y)$ plane

\begin{equation}\label{xycomp}
\vec{A}_j =A_{j,x} \vec{x} + A_{j,y} \vec{y}; \,\, j = 1,2.
\end{equation}
Thus, the components $A_x$ and $A_y$  in (\ref{SYS1}) become

\begin{eqnarray}
\label{INCOND3}  A_x = A_{1,x} + A_{2,x}; \,\,
A_y = A_{1,y} + A_{2,y}.
\end{eqnarray}
Let us  introduce the integral of energy of  $A_x$ and $A_y$
\begin{eqnarray}
p_j = \int\int\int |A_j\left(x,y,z,t \right)|^2 \, dU = const ;
\,\,j=x,y
\end{eqnarray}
where $dU=dxdydz$ and also the integrals of  center of weight in $x$
direction of $A_x$ and $A_y$ are

\begin{eqnarray}
\left\langle X_j \right\rangle = \frac{1}{p_j} \int\int\int
x|A_j\left(x,y,z,t \right)|^2 \, dU;  \,\,j=x,y,
\end{eqnarray}
Only the second derivative $\frac{\partial^2}{\partial x^2}$ in the
system (\ref{SYS1}) is non-commutative operator in regard to $x$
translation. The other differential operators commute with $x$ and
therefore the velocity in $x$ direction of the center of weight  can
be written as

\begin{eqnarray}
\label{VEL}
%\frac{\partial \left\langle x_i \right\rangle}{\partial t}  =
\left\langle  \dot{X_j} \right\rangle = \frac{iv_{gr}}{2 k_0 p_j}
\int\int\int\left[ A_j^{*}\left(x,y,z,t\right)\frac{\partial
A_j\left(x,y,z,t\right)}{\partial x} -
A_j\left(x,y,z,t\right)\frac{\partial
A_j^*\left(x,y,z,t\right)}{\partial x} \right]\, dU;  \,\,j=x,y.
\end{eqnarray}
The acceleration in $x$ direction of the center of weight can be
expressed by the following convolution integral

\begin{eqnarray}
\label{ACCEL} \left\langle  \ddot{X_j}(\Delta x, t) \right\rangle =
\frac{v_{gr} k_0 \tilde{n}_2}{3 p_j} \int\int\int\left[
|A_j\left(x+\Delta x,y,z,t \right)|^2 \frac{\partial }{\partial x}
|A_k\left(x,y,z,t \right)|^2 \right]\, dU;  \,\,j=x,y,
\end{eqnarray}
where $k\neq j$. In the similar way we obtain the expressions of the
accelerations in $y$ and $z$ directions. The total acceleration of
the components $A_x$ and $A_y$  can be presented by the following
vector sums

\begin{eqnarray}
\label{ACCELG}
\left\langle  \ddot{X_j}(\Delta x, t) \right\rangle
\vec{x} + \left\langle \ddot{Y_j}(\Delta y, t) \right\rangle
\vec{y}+ \left\langle \ddot{Z_j}(\Delta z, t) \right\rangle \vec{z}=
\nonumber \\
\frac{v_{gr} k_0 \tilde{n}_2}{3 p_i} \int\int\int \left[
|A_j\left(x+\Delta x,y+\Delta y,z+\Delta z,t \right)|^2 \nabla
|A_k\left(x,y,z,t \right)|^2 \right]\, dU; \,\,j=x,y,
\end{eqnarray}
where with $\nabla=\partial/\partial x+ \partial/\partial
y+\partial/\partial z$ is denoted the gradient operator of the
scalar field $|A_k\left(x,y,z,t \right)|^2 $ and the expression
under the integral is

\begin{eqnarray}
\label{ACCELG1}
|A_j\left(x+\Delta x,y+\Delta y,z+\Delta z,t \right)|^2
\nabla |A_k\left(x,y,z,t \right)|^2 =|A_j\left(x+\Delta x,y,z,t \right)|^2
\frac{\partial (|A_k\left(x,y,z,t \right)|^2 )}{\partial x}+\nonumber\\
|A_j\left(x,y+\Delta y,z,t \right)|^2\frac{\partial
(|A_k\left(x,y,z,t \right)|^2 )}{\partial y}+|A_j\left(x,y,z+\Delta
z,t \right)|^2 \frac{\partial (|A_k\left(x,y,z,t \right)|^2
)}{\partial z}; \,\,j=x,y.
\end{eqnarray}
Here we investigate the simplest case of two
\emph{spherically-symmetric} pulses, located at arbitrary distance
$\Delta x$ in $x$ direction. Then, since $\Delta y=\Delta z=0$, the
convolution integrals in $y$ and $z$ planes are equal to zero.
Therefore we calculate the acceleration in $x$ direction
(\ref{ACCEL}) only.  Substituting the decomposition (\ref{INCOND3})
in (\ref{ACCEL}) we obtain
\begin{eqnarray}
\label{ACCELXA1} \left\langle  \ddot{a}(\Delta x, t)
\right\rangle_{\vec{A}_1} =C_1 \int\int\int
\Bigg[|A_{x_1}\left(x+\Delta x,y,z,t \right)|^2 \frac{\partial
}{\partial x}|A_{y_2}\left(x,y,z,t \right)|^2 +\nonumber\\
+|A_{y_1}\left(x+\Delta x,y,z,t \right)|^2 \frac{\partial }{\partial
x} |A_{x_2}\left(x,y,z,t \right)|^2\Bigg]\, dU,
\end{eqnarray}

\begin{eqnarray}
\label{ACCELXA2} \left\langle  \ddot{a}(\Delta x, t)
\right\rangle_{\vec{A}_2} =C_1 \int\int\int
\Bigg[|A_{x_2}\left(x-\Delta x,y,z,t \right)|^2 \frac{\partial
}{\partial x}
|A_{y_1}\left(x,y,z,t \right)|^2 +\nonumber\\
+|A_{y_2}\left(x-\Delta x,y,z,t \right)|^2 \frac{\partial }{\partial
x} |A_{x_1}\left(x,y,z,t \right)|^2\Bigg]\, dU,
\end{eqnarray}
where by $\left\langle  \ddot{a}(\Delta x, t)
\right\rangle_{\vec{A}_1}$ and $ \left\langle \ddot{a}(\Delta x, t)
\right\rangle_{\vec{A}_2}$ are denoted \emph{the accelerations of
the pulses} (not of the components) with condition $\left\langle
\ddot{a}(\Delta x, t) \right\rangle_{\vec{A}_1} +\left\langle
\ddot{a}(\Delta x, t) \right\rangle_{\vec{A}_2}=0$ and
$C_1=\frac{(p_x+p_y)v_{gr} k_0 \tilde{n}_2}{3 p};\,\,p=p_xp_y$. In
the case of spherically-symmetric functions and circular
polarization $(A_{x_i}=A_{y_i})$ the acceleration of the center
weights can be presented in spherical coordinates

\begin{eqnarray}
\label{ACCELGS}
%\frac{\partial \left\langle x_i \right\rangle}{\partial t}  =
\left\langle  \ddot{a}(\Delta r, t) \right\rangle _{\vec{A}_1}=
 2C_1\int\int\int
\left[ |A_1\left(r+\Delta r,t \right)|^2 \frac{\partial}{\partial
r} |A_2\left(r,t \right)|^2 r^2 \sin \theta \right]\, dr \,d\theta
\, d\varphi,\nonumber\\
\\
\left\langle  \ddot{a}(\Delta r, t) \right\rangle _{\vec{A}_2}=
2C_1\int\int\int \left[ |A_2\left(r+\Delta r,t \right)|^2
\frac{\partial}{\partial r} |A_1\left(r,t \right)|^2 r^2 \sin \theta
\right]\, dr \,d\theta \, d\varphi.\nonumber
\end{eqnarray}
where $\Delta r$ is the distance between the centers of weight of
the pulses. Since the acceleration depends on $\Delta r$
(\ref{ACCELGS}) we can introduce the nonlinear potential

\begin{eqnarray}
\label{POT} V \left( \Delta r,t \right) = V \left( 0,t \right) -
\int_0^{\Delta r} \left\langle  \ddot{a_i}(\Delta r, t) \right\rangle \, \left(
d\Delta r \right).
\end{eqnarray}
Let us suppose that the pulses do not change their shape and
spectrum during the propagation process -- as it can be seen on
Fig.$1b)$ this assumption is correct, when the pulses are at a
sufficient distance from each other. At close distances the
acceleration and potential depend significantly on time. We  use
trial functions with two shapes  $1)$ Gaussian profile (as in the
numerical experiments above)

\begin{eqnarray}
\label{GAUS}
A = A_1 = A_2 = A_0 \exp \left(
-\frac{x^2+y^2+z^2}{2} \right) = A_0 \exp \left( -\frac{r^2}{2}
\right),
\end{eqnarray}
and $2)$ Lorentz profile (such form have the filaments in the
faraway zone of propagation)

\begin{eqnarray}
\label{Lorentz}
 A = A_1 = A_2 = \frac{2 A_0}{1+r^2}.
\end{eqnarray}
\begin{figure}
\begin{center}
\begin{tabular}{c}
\includegraphics[width=140mm,height=55mm]{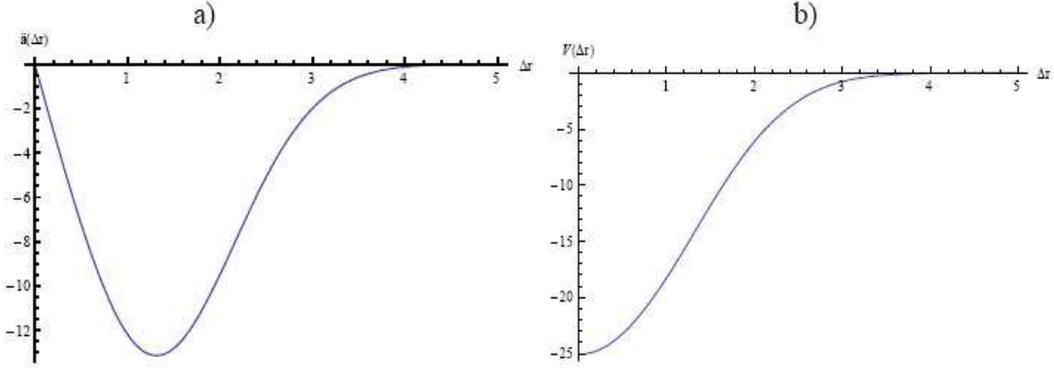}
\end{tabular}
\end{center}
\caption{Graphics of the nonlinear acceleration $\ddot{a}(\Delta r)$
(\ref{ACCELGS}) and potential $V(\Delta r)$ (\ref{POT}) between the
centers of weight in cases of Gaussian pulses (\ref{GAUS}) for
normalized constant $2\pi^2C_1A_0^4=1$.}
\end{figure}
On Fig. 3  the graphics of the nonlinear acceleration $\ddot{a}
(\Delta r)$ (\ref{ACCELGS}) and potential $V(\Delta r)$ (\ref{POT})
between the centers of weight in the cases of Gaussian pulses
(\ref{GAUS}) for normalized constant $2\pi^2C_1A_0^4=1$ are
presented. The same quantities  for  Lorentz pulses (\ref{Lorentz})
are plotted on Fig. 4.  The formulae of the exact solutions of the
convolution integral (\ref{ACCELGS}) and the potentials (\ref{POT})
for the both cases are

\begin{figure}
\begin{center}
\begin{tabular}{c}
\includegraphics[width=140mm,height=55mm]{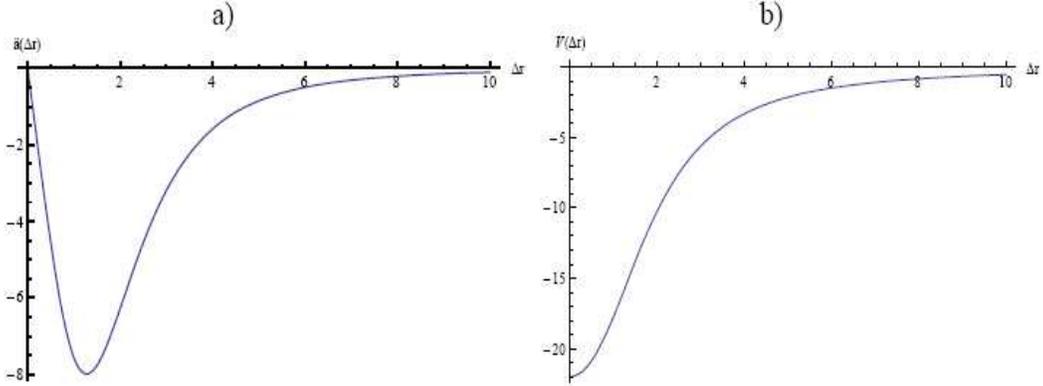}
\end{tabular}
\end{center}
\caption{Graphics of the nonlinear acceleration $\ddot{a}(\Delta r)$
(\ref{ACCELGS}) and potential $V(\Delta r)$ (\ref{POT}) between the
centers of weight in  cases of Lorentz pulses (\ref{Lorentz}). The
potential of Lorentz pulses is approximately twice wider than the
potential of Gaussian ones.}
\end{figure}

\begin{eqnarray}
\label{GAUSA} \left\langle  \ddot{a_i}^{Gaus}(\Delta r)
\right\rangle =-2\sqrt{2\pi} \Delta r \left(3+\Delta
r^2\right)\exp\left(-\Delta r/2\right);\,\, V^{Gaus} \left( \Delta r
\right) = 2\sqrt{2\pi}\left(5+\Delta r^2\right)\exp\left(-\Delta
r/2\right).
 \end{eqnarray}

\begin{eqnarray}
\label{LORENTZ} \left\langle \ddot{a_i}^{Lorentz}(\Delta
r)\right\rangle = -\frac{32\pi\Delta r\left(24+22\Delta r^2+\Delta
r^4\right)}{\left(4+\Delta r^2\right)^4};\,\, V^{Lorentz} \left(
\Delta r \right) = \frac{16\pi\Delta r\left(28+15\Delta r^2+\Delta
r^4\right)}{\left(4+\Delta r^2\right)^3}).
\end{eqnarray}
It is important to be pointed that the potential of Lorentz pulses
is approximately twice wider than the potential of Gaussian pulses.
As a result, the Lorentz type filaments can interact at twice longer
distance than the standard Gaussian type filaments.  In the general
case, \emph{the acceleration and the potential are not stationary}
and as it can be seen from the expressions   (\ref{ACCELGS}) and
(\ref{POT}) depend in addition on the time. That is why the spatial
forms and the spectrums of the pulses at short distances are
modulated significantly. The numerical experiments demonstrate, that
if the pulses are separated along $x$ direction, the forms and the
$k_x$ spectrums of the both pulses become asymmetric.

\section{Conclusions}

We have developed a vector model to describe the processes of
reduction of number of the filaments  as well as the observation of
mergers and  Rogue events during  multi-filament propagation.  It is
known, that in air $P=P_{cr}$ corresponds to intensity of the laser
field of the order of $I\sim 10^{12}$ $W/cm^2$. The main role at
these intensities play the nonlinear $\chi^{(3)}$ effects. The
results from the numerical analysis give confidence for claiming,
that the investigated above processes are result of nonlinear
interactions due to FPPM and CPM mechanisms.   Finally, by using the
method of moments, the merging between spherically-symmetric,
circular polarized filaments is presented as potential interaction.

\section{Acknowledgements}
This work was supported in part by Bulgarian Science Fund under
grant DFNI--I-02/9.

\newpage

\section{List of Figure Captions}

Fig.1 (1a) Energy exchange between two collinear filaments
$\vec{A}_1$ and $\vec{A}_2$ with power slightly above the critical
$P_{cr}$ ($\gamma=1.5$). The pulses are separated initially at
distance $2a=3.4$ and the evolution is governed by the system of
equations (\ref{SYSGAL1}). In the initial conditions (\ref{INCOND1})
the phase difference $\varphi=\pi/4$ correspond to maximal energy
exchange. Due to degenerated FPPM process one of the filaments is
amplified while the other filament enters in linear mode and
vanishes.  (1b) Fusion between the same pulses when they are with
equal initial phases, i. e. $\Delta\varphi=0$. The similar picture
is seen when the FPPM terms are excluded from the equations
(\ref{SYSGAL1}) and only interaction due to CPM is investigated.
With $z_{diff}$ is denoted the diffraction length
$z_{diff}=k_0r_0^2$.

Fig.2 Interaction between three collinear filaments $\vec{A}_1$,
$\vec{A}_2$ and $\vec{A}_3$ governed by the system of equations
(\ref{SYSGAL1}). Two of the pulses  are with equal initial phases,
while the third admits phase difference $\Delta\varphi=\pi/4$  in
respect to others. Fusing between the pulses with equal phases is
seen, while the third one exchanges  energy by FPPM process.

Fig. 3 Graphics of the nonlinear acceleration $\ddot{a}(\Delta r)$
(\ref{ACCELGS}) and potential $V(\Delta r)$ (\ref{POT}) between the
centers of weight in cases of Gaussian pulses (\ref{GAUS}) for
normalized constant $2\pi^2C_1A_0^4=1$.

Fig. 4  Graphics of the nonlinear acceleration $\ddot{a}(\Delta r)$
(\ref{ACCELGS}) and potential $V(\Delta r)$ (\ref{POT}) between the
centers of weight in  cases of Lorentz pulses (\ref{Lorentz}). The
potential of Lorentz pulses is approximately twice wider than the
potential of Gaussian ones.

\end{document}